\documentstyle[12pt,epsf,epsfig,a4]{article}
\begin{document}
\begin {titlepage}
\begin{flushleft}
FSUJ TPI QO-2/98
\end{flushleft}
\begin{flushright}
February, 1998
\end{flushright}
\vspace{20mm}
\begin{center}
{\Large {\bf
Homodyne measurement of exponential phase moments 
for quantum-phase reconstruction}\\[3ex]
\Large M. Dakna $^{\rm a}$, G. Breitenbach $^{\rm b}$,
J. Mlynek $^{\rm b}$,\\
T. Opatrn\'{y} $^{\rm c}$, S. Schiller $^{\rm b}$, D.--G. Welsch $^{\rm a}$}\\[2.ex] 
$^{\rm a}$ Friedrich-Schiller-Universit\"at  Jena,
Theoretisch-Physikalisches Institut,
Max-Wien-Platz 1, 07743 Jena, Germany\\[1ex]
$^{\rm b}$ Fakult\"at f\"ur Physik, Universit\"at Konstanz,
D-78457 Konstanz, Germany\\[1ex]
$^{\rm c}$ Palack\'{y} University, 
Department of Theoretical Physics,\\
Svobody 26, 77146 Olomouc, Czech Republic 

\vspace{25mm}
\end{center}
\begin{center}
\bf{Abstract}
\end{center}
We directly sample the exponential moments of the canonical phase for 
various quantum states from the homodyne output. 
The method enables us to study the phase properties
experimentally, without making the detour via reconstructing
the density matrix or the Wigner function and calculating
the phase statistics from them. In particular,
combing the measurement with a measurement of the photon-number 
variance, we verify fundamental number--phase uncertainty.   
\end{titlepage}


Optical homodyne tomography has been a powerful method
for quan\-tum-state measurement, because the measured
quadrature-component distributions $p(x,\vartheta)$ $\!=$ 
$\langle x,\vartheta|\hat{\varrho} |x,\vartheta\rangle$ 
contain all the information about the quantum state 
[$|x,\vartheta\rangle$ being the eigenstates of
$\hat{x}(\vartheta)$ $\!=$ $\!2^{-1/2}$ $\!(e^{-i\vartheta}\hat{a}$
$\!+$ $\!e^{i\vartheta}\hat{a}^{\dagger})$]. In this way,
the Wigner function \cite{VogelRisken} and 
the density matrix \cite{Ariano,Schiller,Breitenbach}
can be inferred from experimental data. 
When the state (in a chosen representation) is known,
then the mean values of arbitrary quantities can be calculated.

Each measurement is associated with errors
(at least, the statistical error) which propagate in the 
calculation procedure. Hence, the error of a calculated quantity  
can be too large to be acceptable. For example, if we are interested 
in photon-number moments $\langle\hat{n}^{k}\rangle$, 
we have to multiply the diagonal density-matrix elements $\varrho_{n,n}$ 
by $n^{k}$ and perform the sum. Since the statistical error of  
$\varrho_{n,n}$ does not vanish with increasing $n$,
the total error is infinite after summing all the terms. 
Truncation of the sum can avoid this trouble for
sufficiently low-order moments, whereas for high-order moments
also small values of $\varrho_{n,n}$ with large
$n$ may be essential. To overcome this problem, it has been suggested 
to sample the desired quantities directly from the measured data, without 
the detour via the density matrix (or other state representations).
Formulas have been derived that are suited
for direct sampling of normally ordered moments of photon
creation and destruction operators \cite{Richter2}, 
and an extension to arbitrary 
quantities that admit normal-order expansion 
has been given \cite{DAriano}.
Quite recently sampling functions for measuring the exponential
moments of the canonical phase have been derived \cite{Dakna1}.

In this paper we apply the sampling method to an experimental 
determination of the exponential moments of the canonical 
phase for various single-mode quantum states. The results are then 
used for determining the phase distribution as Fourier transform of
the exponential phase moments. In particular, the sampled 
first exponential phase moment already defines a phase uncertainty 
\cite{Bandilla,Holevo}. Using the (simultaneously) sampled
photon-number variance a verification of the number--phase uncertainty  
predicted theoretically in \cite{Holevo} is given.



The exponential phase moments $\Psi_k$ are defined by the Fourier 
components of the phase distribution $P(\varphi)$, i.e.,
$\Psi_k$ $\!=$ $\!\int_{2\pi} d\varphi$ $\!e^{ik\varphi} 
P(\varphi)$. For the canonical phase they are given by \cite{Car}
$\Psi_{k}$ $\!=$ $\!\langle \hat E^{k} \rangle$
if $k$ $\!>$ $\!0$, and $\Psi_{k}$ $\!=$ $\!\Psi_{-k}^{*}$ 
if $k$ $\!<$ $\!0$, where
$\hat E$ $\!=$ $\!(\hat n$ $\! +$ $\!1)^{-1/2} \hat a$,
$\hat n$ $\!=$ $\!\hat a^{\dagger} \hat a$ being 
the photon-number operator. It can be shown that $\Psi_{k}$ 
can be obtained from the quadrature-component distributions
$p(x,\vartheta)$ as
\cite{Dakna1}
\begin{eqnarray}
\label{c6}
\Psi_k =\int_{2\pi} d \vartheta \int_{-\infty}^{\infty} dx
\, K_{k}(x,\vartheta)\,p(x,\vartheta),
\end{eqnarray}
where $K_{k}(x,\vartheta)$ is a well-behaved 
integral kernel suited for direct sampling of $\Psi_{k}$
from the homodyne output for any normalizable quantum state.

Knowing $\Psi_k$, the phase distribution $P(\varphi)$
can be obtained according to
$P(\varphi)$ $=$ $(2\pi)^{-1}$ $\sum_{k=-\infty}^{\infty}$
$e^{-ik\varphi} \Psi_k$.
However, the first moment already contains essential information 
about the phase properties. It can be used to introduce a mean phase
$\bar \varphi$ $\!=$ $\!\arg \Psi _{1}$ and a phase uncertainty 
$\Delta\varphi$ $\!=$ $\!\arccos | \Psi _{1}|$ ,
which implies a number--phase uncertainty relation \cite{Holevo}
\begin{equation}
\label{ur}
\Delta n \tan \Delta \varphi \ge {\textstyle\frac{1}{2}} . 
\end{equation}
Note that for the number-uncertainty
$\Delta n$ $\!=$ $\!( \langle\hat{n}^{2}\rangle$ 
$\!-$ $\!\langle\hat{n}\rangle^{2} )^{1/2}$
the quantities $\langle \hat n \rangle$ and $\langle \hat n^{2} \rangle$
can also be obtained by direct sampling according to a relation of
the form (\ref{c6}), with the integral kernel being given in
\cite{Richter2}. Hence,
homodyne detection can be regarded as the most direct way
that has been known so far for experimental verification of the 
uncertainty relation (\ref{ur}).



The experimental setup 
is the same as in \cite{Breitenbach}.  
Its central unit is a monolithic standing-wave lithium-niobate
optical parametric amplifier (OPA) pumped by a  
frequency-doubled continuous-wave Nd:YAG laser ($1064$ nm). 
Operated below threshold, the OPA is a source of squeezed vacuum.
We study the spectral components of the field around a frequency  
offset by $\Omega/2\pi$ $\!=$ $\!1.5$ or $2.5$ MHz from the optical 
frequency $\omega$, to avoid low-frequency laser excess noise.
To generate bright light, a very weak wave split off the main laser 
beam is phase-modulated by an electro-optic modulator (EOM) at the 
frequency $\Omega$ and injected into the OPA through its high reflector 
port. The carrier frequency $\omega$ is kept on-resonance with the  
cavity and the two ``bright'' sidebands $\omega\pm\Omega$ are well
within the cavity bandwidth. By turning the modulation off, we
obtain squeezed vacuum, by blocking the OPA pump
wave, we are left with coherent excitations.

The signal is analyzed at a  homodyne detector, whose output  
current is mixed with an electrical local oscillator phase-locked to  
the modulation frequency, low-pass filtered and recorded  with a  
high speed A/D converter.
Since the squeezed states are  
essentially two-mode states, a two-mode detection is crucial for  
obtaining the correct statistics of the light field. We remark  
however, that this type of measurement may need modifications for  
general states of the light field.
The quadrature-component distributions 
$p(x,\vartheta)$ are obtained by subdividing the recorded noise 
traces into 128 equal length intervals and subsequently forming histograms
of 256 amplitude bins, normalizing the absolute bin width using as  
reference the distribution of a vacuum state.



\begin{figure}[htp]
\centering\epsfig{figure=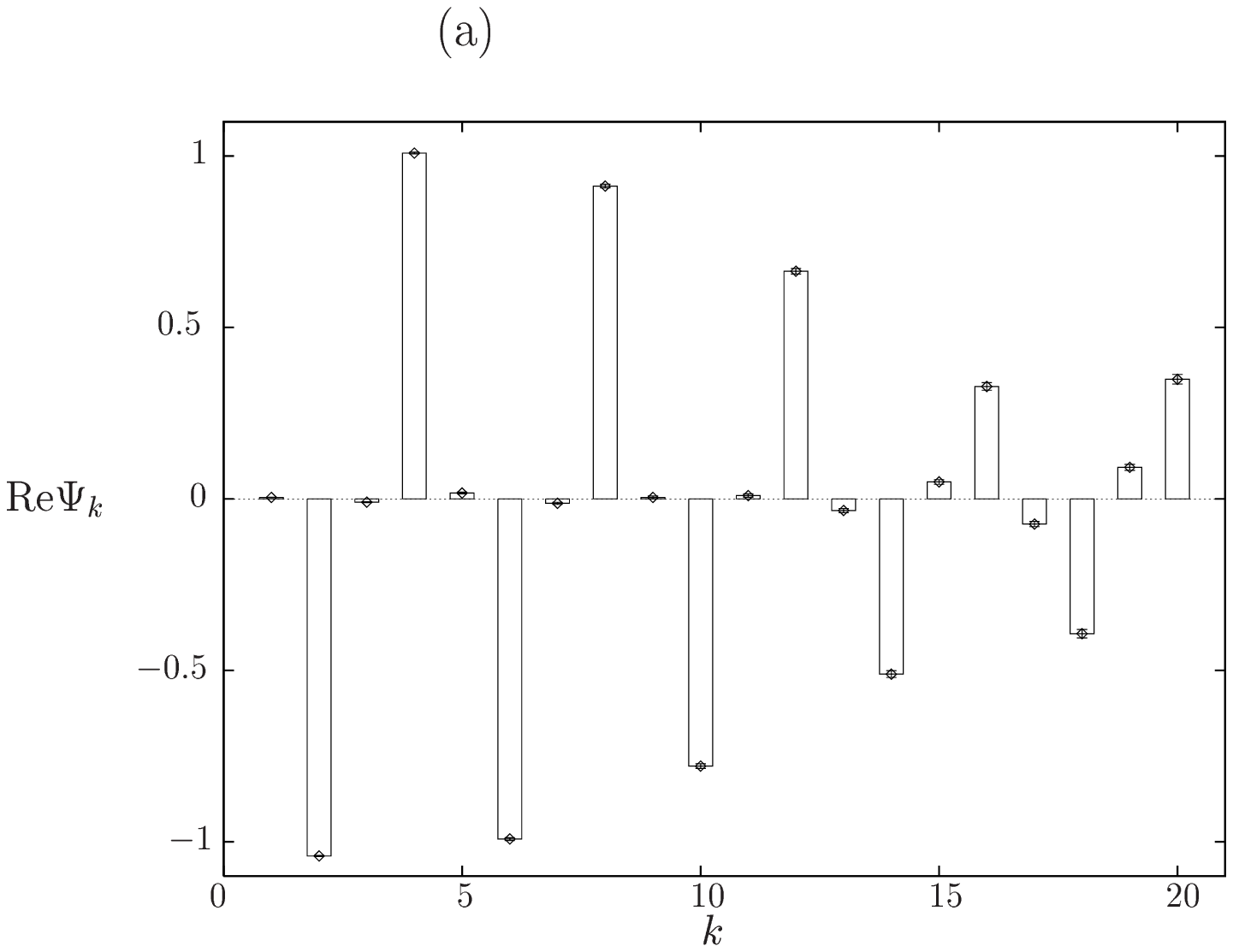,width=0.8\linewidth}

~
\vspace{0.1cm}

~
\centering\epsfig{figure=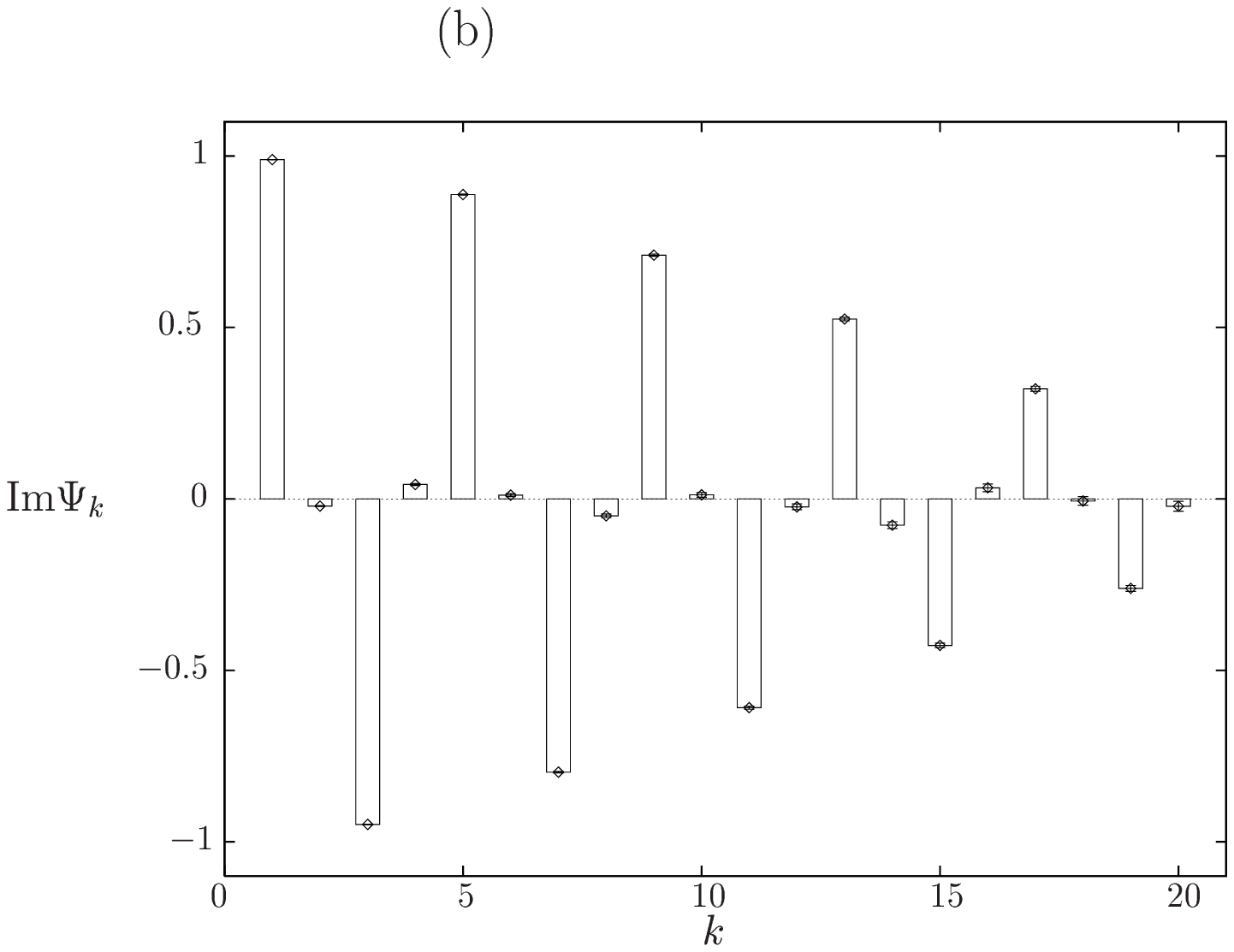,width=0.8\linewidth}
\caption{
Measured exponential phase moments $\Psi_{k}$ for a phase squeezed 
state [state (F) in Tab.~\protect\ref{Tab1}] [(a) real part; 
(b) imaginary part]. 
\label{F2}
}
\end{figure}

\begin{figure}[htp]
\centering\epsfig{figure=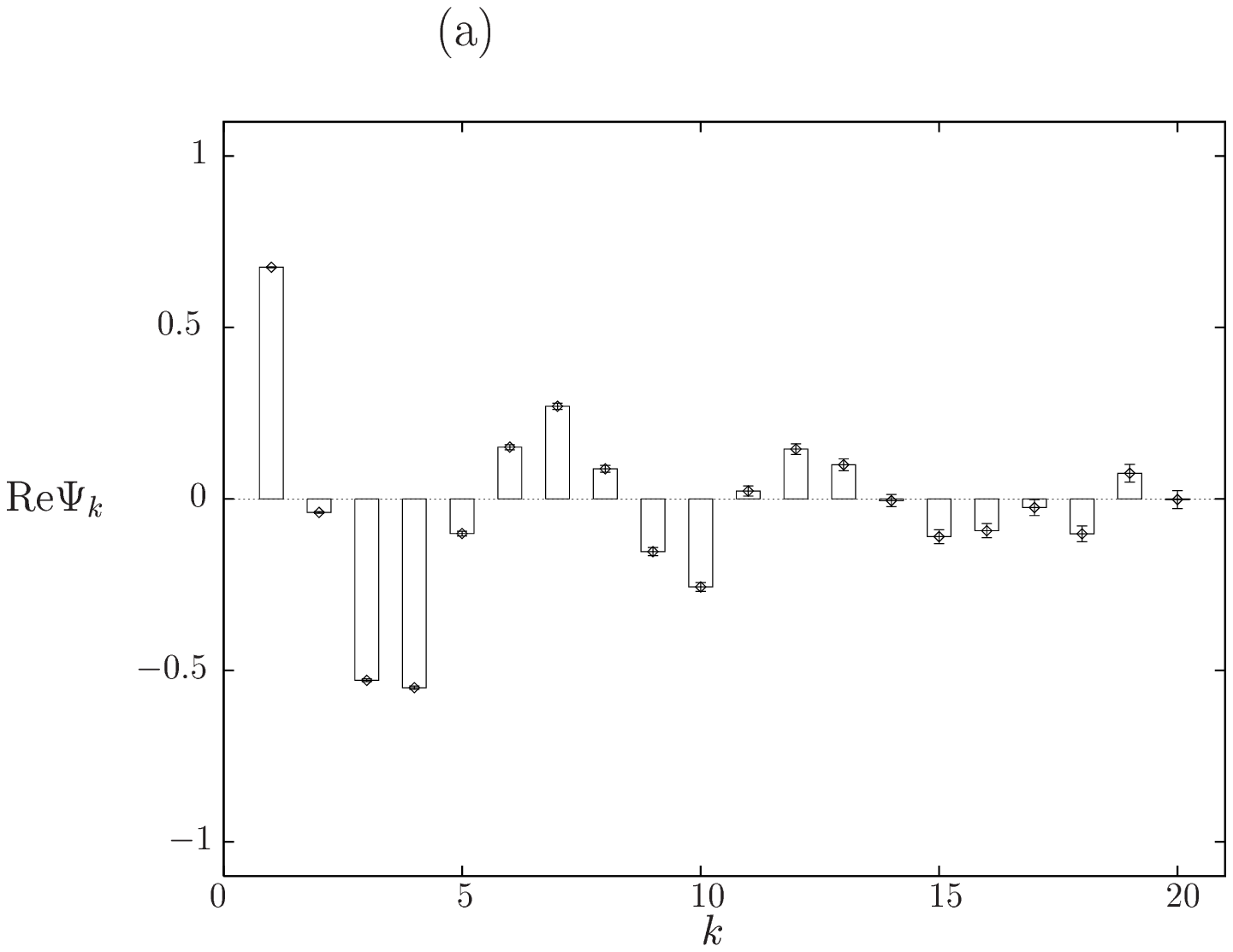,width=0.8\linewidth}

~
\vspace{0.1cm}

~
\centering\epsfig{figure=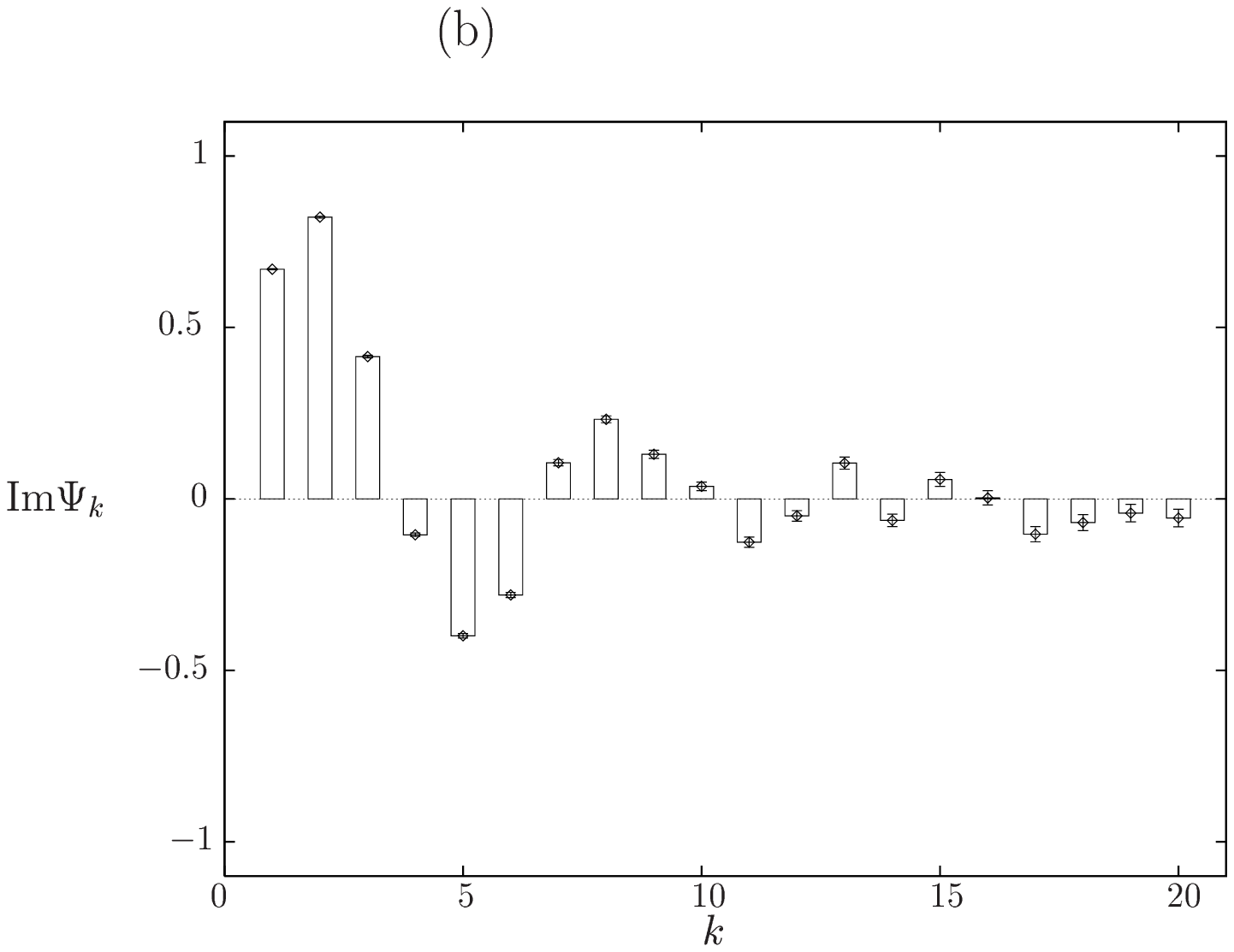,width=0.8\linewidth}
\caption{
Measured exponential phase moments $\Psi_{k}$ for
a state squeezed at a phase angle of $48^{\circ}$
[state (H) in Tab.~\protect\ref{Tab1}]
[(a) real part; (b) imaginary part].  
\label{F3}
}
\end{figure}

\begin{figure}[htp]
\centering\epsfig{figure=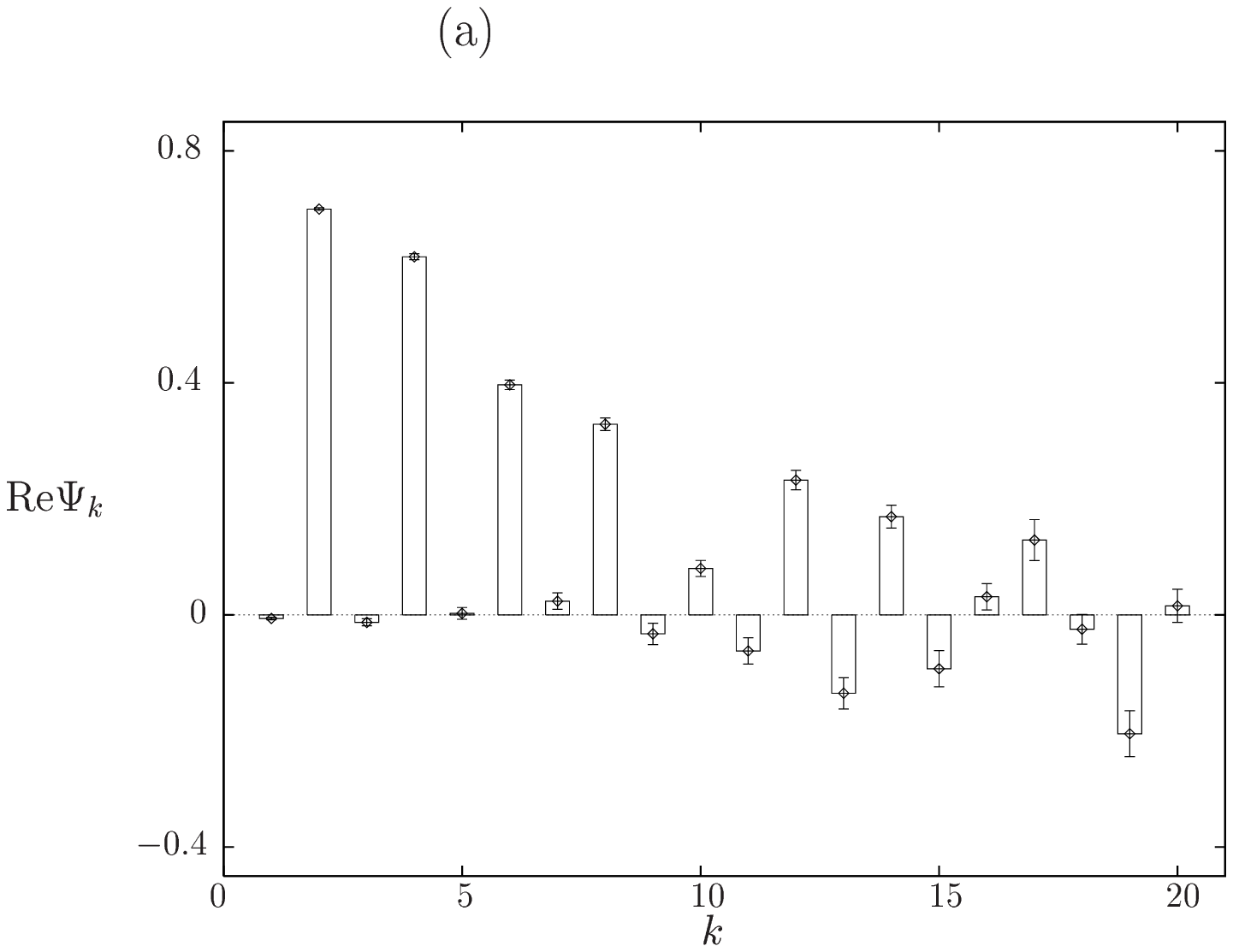,width=0.8\linewidth}

~
\vspace{0.1cm}

~
\centering\epsfig{figure=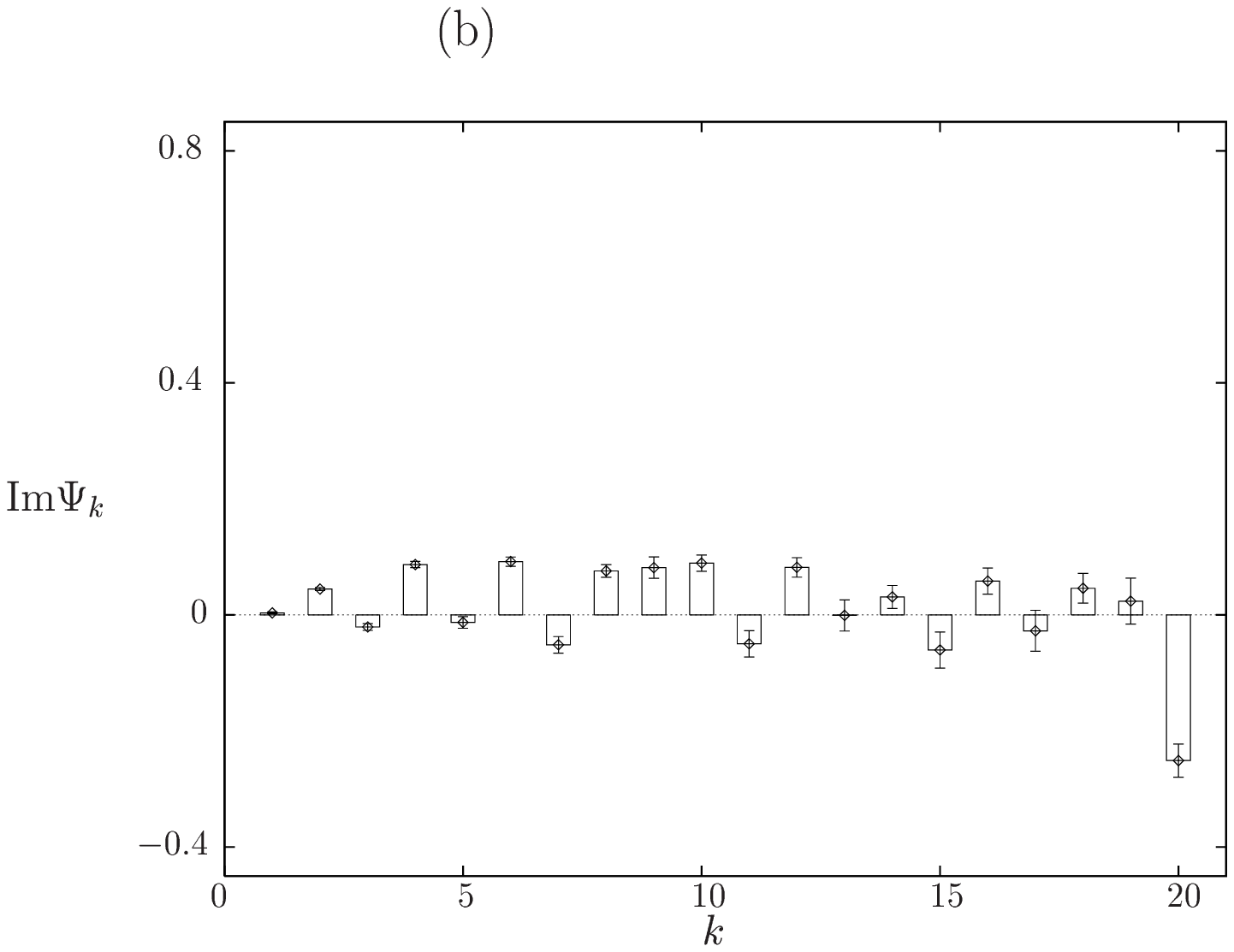,width=0.8\linewidth}
\caption{
Measured exponential phase moments $\Psi_{k}$ for
a squeezed vacuum [state (G) in Tab.~\protect\ref{Tab1}]
[(a) real part; (b) imaginary part].  
\label{F4}
}
\end{figure}

\begin{figure}[htp]
\centering\epsfig{figure=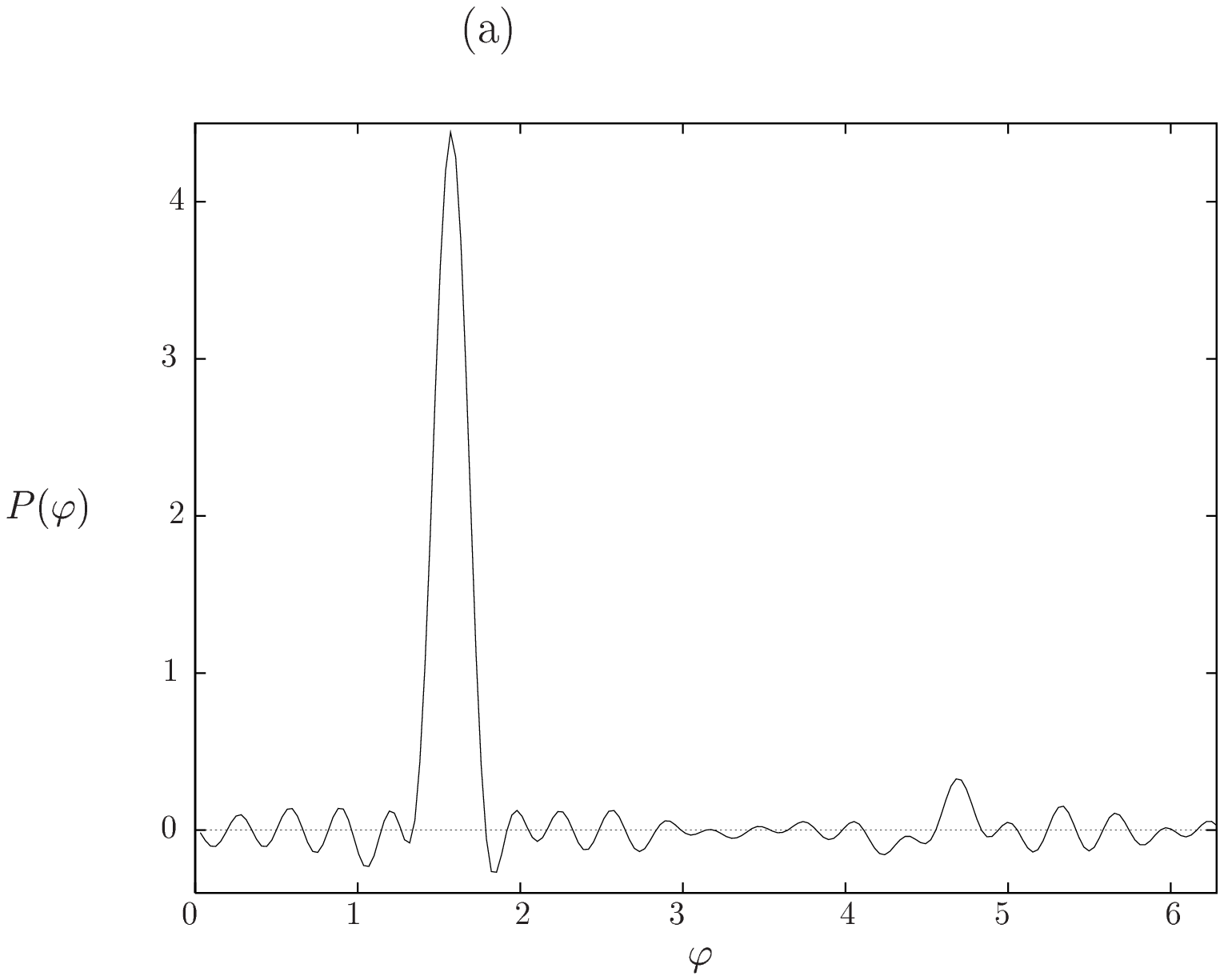,width=0.6\linewidth}

~
\vspace{0.05cm}
~
\centering\epsfig{figure=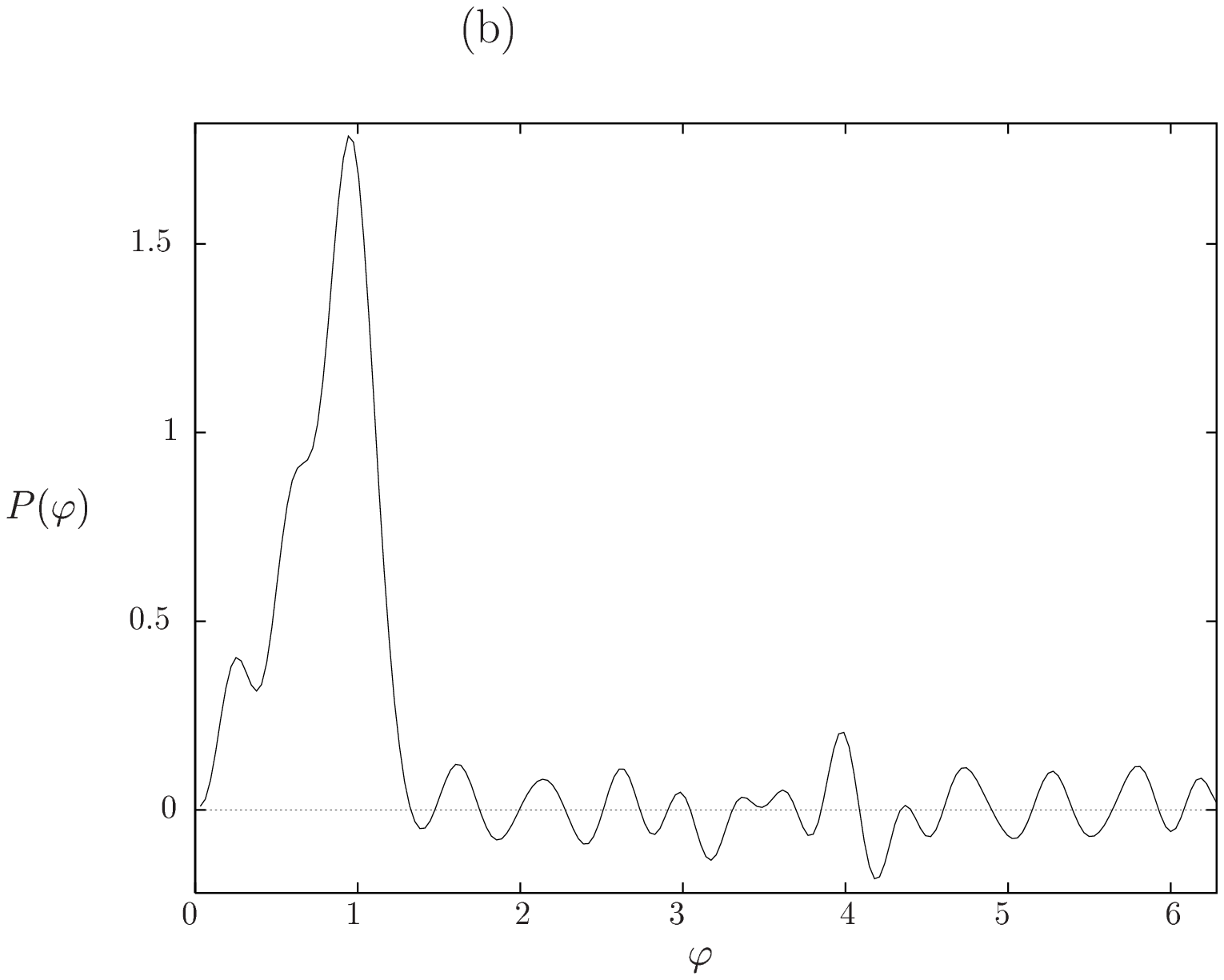,width=0.6\linewidth}
~
\vspace{0.05cm}
~
\centering\epsfig{figure=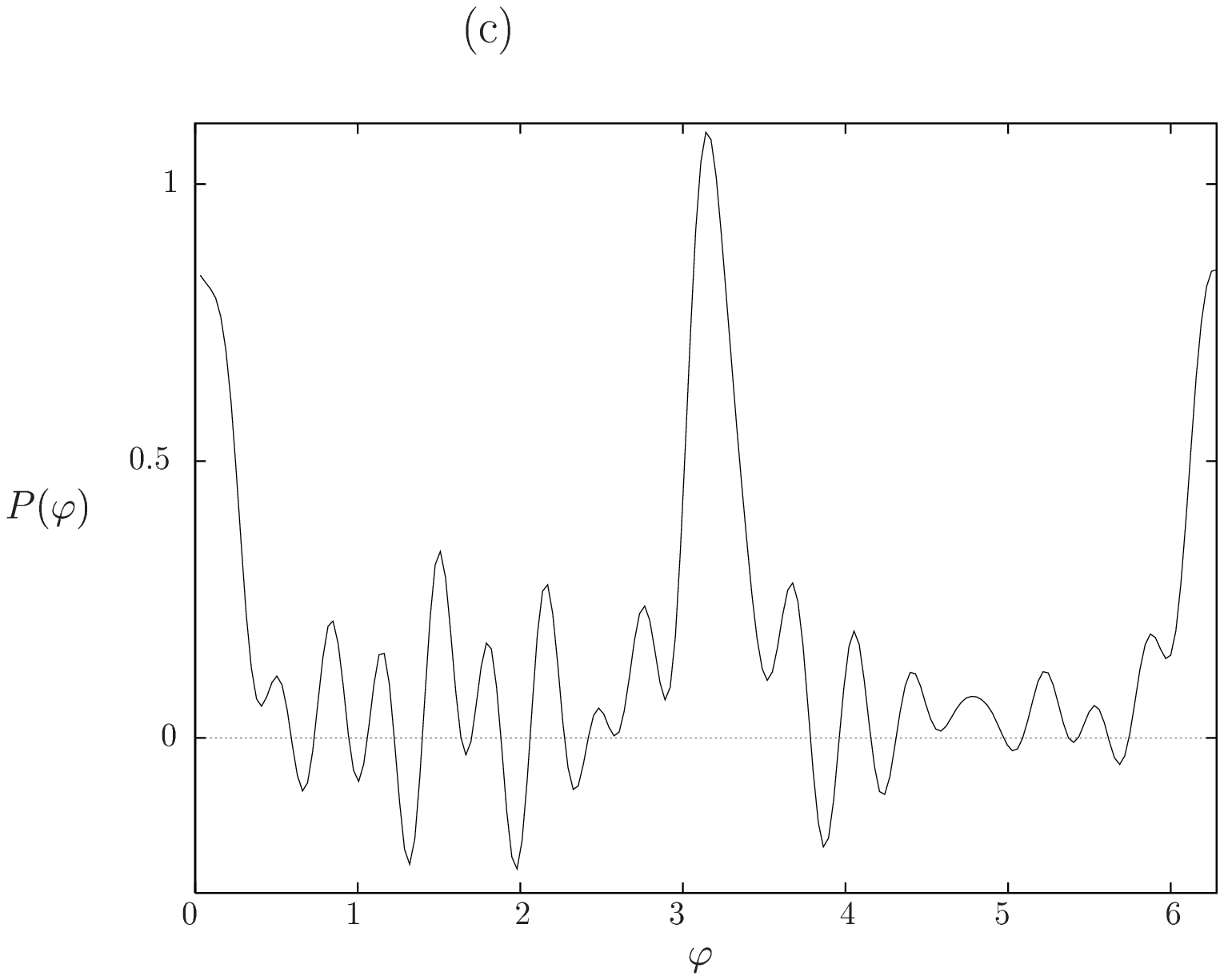,width=0.6\linewidth}
\caption{The canonical phase distribution $P(\varphi)$ reconstructed from
$\!20$ measured exponential phase moments 
$\Psi_{k}$ given in Figs.~\protect\ref{F2} -- \protect\ref{F4} is shown
for (a) a phase-squeezed state [state (F) in
Tab.~\protect\ref{Tab1}], (b) for 
a state squeezed at a phase angle of $48^{\circ}$
[state (H) in Tab.~\protect\ref{Tab1}] and (c) for a squeezed 
vacuum [state (G) in Tab.~\protect\ref{Tab1}].
\label{F5}
}
\end{figure}

In Figs.~\ref{F2}--\ref{F4}, the sampled exponential phase
moments $\Psi_{k}$, $k$ $\!=$ $\!1,2,\dots,20$, are shown for
a phase-squeezed state (Fig.~\ref{F2}), 
a state squeezed at a phase angle of $48^{\circ}$
(the difference between the argument of the squeezing parameter
and the argument of the displacement parameter)
and a squeezed vacuum (Fig.~\ref{F4}).  
The error bars indicate the statistical error. 
Since the main source of inaccuracy is the fluctuation
of the local oscillator, the error is dominated by the systematic one.
The canonical phase distributions obtained from the sampled moments 
(Fourier components) are plotted in Fig.~\ref{F5}. 
Since in Fig.~\ref{F2} $|\Psi_{k}|$ decreases with increasing 
$|k|$ slower than in Fig.~\ref{F2}, the phase distribution of the 
phase-squeezed state in Fig.~\ref{F5}(a) is more sharply peaked
than that in Fig.~\ref{F5}(b). Figure.~\ref{F5}(c) clearly reveals
the double-peak structure of the phase distribution of a squeezed vacuum.
Note that the small oscillations in the figures 
(which also include negative values) mainly result from 
systematic errors.

Examples of $\bar \varphi$ and $\Delta \varphi$ together with the 
measured mean photon number $\bar n$ $\!=$ $\!\langle\hat{n}\rangle$ 
and photon-number uncertainty $\Delta n$ are given in Tab.~\ref{Tab1} 
for various states prepared in the experiment. 
The last row shows the resulting values of the number--phase
uncertainty product $\Delta n \tan \Delta \varphi$,
which are in full agreement with the predicted inequality (\ref{ur}).
The (near-)coherent states (A,B) and phase-squeezed states (E,F) are 
seen to exhibit relatively small phase uncertainties. Note that the 
coherent state (B) has the smallest phase uncertainty. Relatively large 
phase uncertainties are observed for the amplitude-squeezed states (C,D) 
and the 
state (H) squeezed at a phase angle of $48^{\circ}$.
As expected, the 
near-maximum phase uncertainty $\Delta\varphi$ $\!\approx$ $\!\pi/2$ 
corresponds to the squeezed vacuum (G) (the ``ellipse'' in the phase 
space is centred at the origin). Therefore, for this state the 
uncertainty product $\Delta n$ tan$\Delta \varphi$
achieves a very large value. The smallest value of the uncertainty product 
is observed for the coherent state (A). It is close to its limit
1/2. With respect to the photon number, the amplitude-squeezed 
state (C) is seen to be sub-Poissonian. 



In summary, we have sampled the exponential moments of the
canonical phase directly from the homodyne output for various coherent
and squeezed states produced in a continuous optical field by 
means of parametric amplification. This has enabled us to
study the canonical phase statistics experimentally, without 
the necessity of state reconstruction, which saves calculation effort
and reduces the statistical error. 
In particular, from the sampled first-order exponential phase moment 
and the simultaneously sampled first- and second order photon-number
moments we have determined phase and number uncertainties
and shown that the uncertainty products are
in agreement with the theoretical prediction.


This work was supported by the Deutsche Forschungsgemeinschaft.

\begin{table}
\centering\epsfig{figure=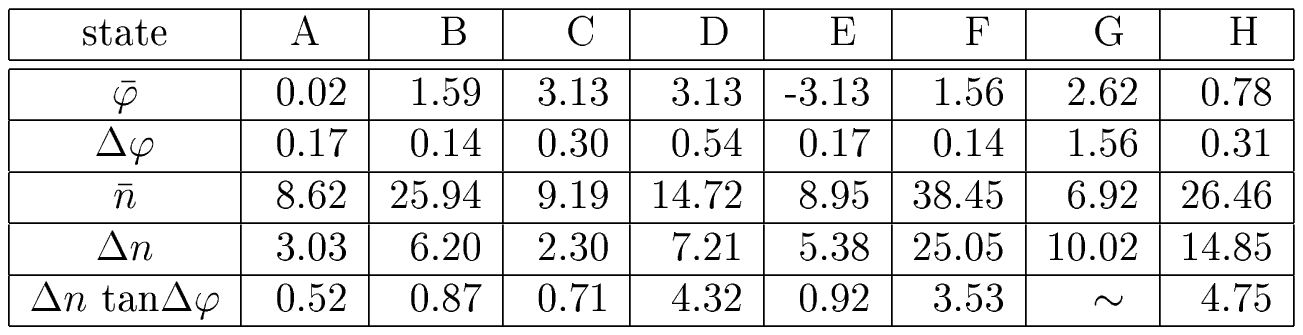}

\vspace{2cm}

\caption{
Measured values of $\bar \varphi$, $\Delta \varphi$,
$\bar n$ $\!=$ $\!\langle\hat{n}\rangle$, and 
$\Delta n$, and the resulting values of the 
number--phase uncertainty product $\Delta n \tan \Delta \varphi$
for various quantum states [(A,B) coherent states;
(C,D) amplitude-squeezed states; (E,F), phase-squeezed states; 
(G) squeezed vacuum; 
(H) state squeezed at a phase angle of $48^{\circ}$].
\label{Tab1}
}
\end{table}
\newpage
\newpage
\newpage

\end{document}